\documentclass[usenatbib,usegraphicx,useAMS]{mn2e}
\usepackage{mnoverride}
\usepackage{setspace}
\usepackage{soul}
\usepackage[utf8]{inputenc}
\usepackage[T1]{fontenc}
\usepackage[fleqn]{amsmath}
\usepackage{amssymb}
\usepackage{mathptmx} 

\usepackage[usenames,dvipsnames]{color}
\DeclareMathSymbol{\utau}{0}{UPM}{"1C}
\DeclareMathAlphabet{\mathcal}{OMS}{cmsy}{m}{n}
\newcommand{\Rmax}{R_{\rm max}}

\newif\ifusehyperref
\usehyperreftrue 
\ifusehyperref
 \usepackage[unicode,breaklinks]{hyperref}
 \usepackage[all]{hypcap}
 \hypersetup{backref,
             bookmarks,
             dvips,
             pdftitle=A conservation-based method for simulating the inspiral of binary black holes,
             pdfsubject=Astrophysics \& Cosmology,
             pdfkeywords={black hole physics -- stars: kinematics and dynamics -- galaxies: nuclei},
             pdfdisplaydoctitle,
             colorlinks=true,
             citecolor=Blue,
             pdfauthor=Yohai Meiron,
             unicode}
\else
 \newcommand{\href}[2]{#2}
\fi

\title[A conservation-based method for simulating BBHs]{A conservation-based method for simulating the inspiral of binary black~holes}
\author[Y. Meiron and A. Laor]{Yohai Meiron\thanks{E-mail: \href{mailto:ym@physics.technion.ac.il}{ym@physics.technion.ac.il} (YM); \newline\href{mailto:laor@physics.technion.ac.il}{laor@physics.technion.ac.il} (AL)}
and Ari Laor\footnotemark[1]\\
Department of Physics, Technion -- Israel Institute of Technology, Haifa 32000, Israel}
\begin{document}

\date{Accepted 2012 January 17. Received 2012 January 16; in original form 2011 October 28}

\pagerange{\pageref{firstpage}--\pageref{lastpage}} \pubyear{2010}

\maketitle

\label{firstpage}

\begin{abstract}
We present a new approach to studying the evolution of massive black hole binaries in a stellar environment. By imposing conservation of total energy and angular momentum in scattering experiments, we find the dissipation forces that are exerted on the black holes by the stars, and thus obtain the decaying path of the binary from the classical dynamical friction regime down to subparsec scales. Our scheme lies between scattering experiments and $N$-body simulations. While still resolving collisions between stars and black holes, it is fast enough and allows to use a large enough number of particles to reach a smooth and convergent result. We studied both an equal mass and a 10:1 mass ratio binaries under various initial conditions. We show that while an equal mass binary stalls at a nearly circular orbit, a runaway growth of eccentricity occurs in the unequal mass case. This effect reduces the timescale for black hole coalescence through gravitational radiation to well below the Hubble time, even in spherical and gasless systems formed by dry mergers.
\end{abstract}


\purplethingy{Accepted}

\section{Introduction}
Supermassive black hole (BH) binaries are formed as a result of galactic mergers. The two BHs sink towards the bottom of the potential well and form a bound pair at the centre of the stellar distribution. According to the classical picture \citep*{begelman80} the binary hardens until the loss cone is depleted (i.e. depleted of stars on low angular momentum orbits that pass close enough to the binary for a significant interaction), and stalls at a separation too large for the emission of gravitational radiation to cause rapid inspiral and coalescence (the `final parsec problem'). This picture has been tested extensively using $N$-body simulations and scattering experiments. Early numerical studies concentrated on the simple case of an equal mass binary on a circular orbit within a spherically symmetric stellar distribution. Those studies (e.g. \citealt{makino97}; \citealt{qh97}; \citealt{mm01}; \citeyear{mm03}), with a relatively small number of particles, were somewhat inconsistent with each other and could not confirm loss cone depletion due to spurious numerical relaxation over long time scales.

In recent years, more detailed investigations have been performed. Apart from the increased number of particles and longer integration times, various complications have been considered such as triaxiality and rotation (e.g. \citealt{berczik06}; \citealt*{khan11}), post Newtonian corrections (e.g. \citealt{berentzen09}) massive perturbers (e.g. \citealt{hagai08}) and gas discs (e.g. \citealt{jorge09}). These studies found that binary coalescence in less than the Hubble time is feasible even if only one of these factors is present. Thus, the final parsec is potentially not a problem in a realistic merger remnant.

More recent work (e.g. \citealt{iwasawa11}; \citealt*{sesana11}) has focused on the evolution of the binary eccentricity rather than just the binary separation. For high mass ratios ($\sim$ 100:1), these authors showed that even in a spherically symmetric environment, eccentricity can increase to almost unity. Very high eccentricity means that in pericentre the two BHs can be close enough together that gravitational wave emission becomes efficient. This enhanced energy loss to gravitational radiation (as compared to the circular case with the same semi-major axis) can reduce coalescence timescale to well below the Hubble time.

Direct $N$-body simulations are the most accurate way to study binary BH (BBH) evolution, but they are computationally expensive. It is therefore difficult to perform diverse enough tests to cover the problem's parameter space. By compromising for an unrealistically small number of particles, one introduces spurious relaxation. In spherical galaxies, this process drives unrealistic loss cone repopulation (\citealt{yu02}; \citealt{mm03}) and causes the hardening rate to be highly $N$-dependent (\citealt{makino04}; \citealt*{berczik05}). \citet{berczik06} followed the evolution of a BBH in triaxial and rotating galaxy models and found that the hardening rate was $N$-independent, implying a collisionless mode of loss cone repopulation.

Our previous work dealt with a possible stellar kinematical signature of stars around a BBH. In \citet{meiron10} we produced kinematical maps (projected maps of mean stellar velocity, velocity dispersion and higher velocity moments); this was achieved by scattering stars on a BBH on a fixed circular orbit. To produce more realistic kinematical maps, we looked for a way to calculate a BBH path fast and accurately so it could be used for following scattering experiments. Good kinematical maps require a very large number ($N\lesssim 10^8$) of stars on small scales, otherwise the high moments of the line of sight velocity distribution are poorly resolved. $N$-body simulations cannot be made yet with such a large number of particles on small enough scales, but this is not a problem for scattering experiments which are performed on a precalculated BBH path. Since we were already working with scattering experiments, we derived a way to use the existing code base and adapt it to work more like an actual $N$-body simulation, where the effect of scatterings on the BBH orbit is taken into account, giving the orbital evolution.

In this paper we present a new scheme we developed to simulate the inspiral of the BBH from the galactic scale to the hard binary scale. Our method is based on imposing conservation of total energy and angular momentum, instead of directly summing the forces of individual stars on the BHs, and lies between scattering experiments and $N$-body simulations. Studying this type of systems with scattering experiments is significantly faster than a full $N$-body treatment and is appropriate when the bulge has relaxed. The phase space stellar distribution is followed accurately and a more realistic number of stars can be included. Thus, we are able to run many simulations and probe a large range of parameters under reasonable physical assumptions, as well as the convergence of the solution, and compare the inspiral timescale, the stalling radius, and the eccentricity evolution to earlier calculations. Since in our method we obtain the forces on the BHs, it is also possible to test analytical expressions for dynamical friction. The original treatment by \citet{chandra} is for a homogeneous background, therefore we compare our results to \citet{just11} who investigate dynamical friction in power law cusps. Our solution extends beyond the range of validity of their formula and into the hard binary regime of the BHs.

We studied the evolution of both an equal mass binary and a 10:1 mass ratio binary under different initial conditions. We found that both cases presented stalling of the semi-major axis, but in the unequal mass case, eccentricity tended to grow towards unity on timescales well below the Hubble time.

In Section \ref{sec:math} we give a general mathematical formulation of our simulation scheme. The model for the specific simulations we performed in this work is described in Section \ref{sec:model} while a technical description of the scheme appears in Section \ref{sec:algorithm}. In Section \ref{sec:results} we present the results of all BBH simulations and discuss their physical implication. In Section \ref{sec:df} we discuss dynamical friction, compare to earlier results, and show that our code is applicable in very large radii, thus potentially helpful in future studies of phenomena related to dynamical friction. Finally, we give a short summary in Section \ref{sec:summary}.

\section{Mathematical Formulation}\label{sec:math}
\subsection{Motivation}
The two basic ideas of our solution are the separation of timescales and the balance of energy and angular momentum between the BHs and the stars. There are three timescales in the system corresponding to changes in:
\begin{enumerate}
\item stellar orbits following close encounters
\item the BHs' orbits
\item the background stellar potential
\end{enumerate}
In more detail: a star significantly changes its original orbit during a close encounter with a BH, but the BH's path is only slightly perturbed as the force exerted on it is due to many small `scattering' events; the change of the background stellar potential is due to the collective response of stars to the perturbation, which evolves on the dynamical timescale. A yet longer timescale would be of 2-body relaxation, which is longer by a factor of $\sim 0.1 N/\ln N$ \citep{bt} and is expected to be well above the Hubble time for a real galaxy core. However, relaxation time is potentially significantly shorter if the dominant relaxation mode is not 2-body relaxation.

The most basic kind of an $N$-body simulation uses very small steps, after each the vector forces exerted by all the field stars (particles) are summed up and applied to the BHs in the next step; each particle is propagated the same way. Energy and angular momentum are globally conserved (within a given numerical error tolerance) since it is a closed system. If $N$ is not large enough, noisy potential and unrealistically massive stars lead to spurious relaxation. If the time step is too large, close encounters cannot be resolved.

Our scheme uses a `large' $\Delta t$, which we call an {\it interval}, within which there is one or more actual {\it steps} of the ODE solver. Let BH number $i$ (where $i\in\{1,2\}$) propagate one interval between times $t_0$ and $t_1=t_0+\Delta t$ from some vector position ${\bf s}_i(t_0)$ to ${\bf s}_i(t_1)$; we use the term {\it segment} for the path length. After each interval, the energies and angular momenta of all the particles are summed up. If no work was done and no torque exerted due to the background stellar potential, then the energy $\Delta E_\star$ (angular momentum $\Delta L_\star$) gained by the stars during some interval must be equal and opposite to the energy $\Delta E_{{\rm BH}}$ (angular momentum $\Delta L_{{\rm BH}}$) lost to the BHs during this interval. For simplicity, we use a static and spherically symmetric model for the stellar potential, so these requirements are automatically fulfilled. The scheme can accommodate more complicated models as well: the demand for spherical (or more generally, axial) symmetry, which is necessary if one assumes that change of a star's angular momentum is only due to interaction with the BHs, can be relaxed if  the torque component due to the stellar bulge is considered separately; a slowly varying potential can also be taken into account if the proper adjustments are made to the code.

The evolution of a BBH is dominated by close and fast encounters with the field stars and the evolution of the stellar gravity field is only a secular effect. Thus, the part of $\Delta E_\star$ due only to the change in stellar potential is negligible after each interval, justifying the use of a static model. However, at late times, the initial stellar potential is no longer consistent with the actual spatial distribution. The static potential assumption is still reasonable as far as the BHs are concerned, since by the time that any significant evolution of the background potential has taken place, the BHs will have fallen deep enough in the potential well, where the dominant force is the other BH's gravity rather than the background stellar potential (so that the exact shape of the potential well does not matter). Stars further out orbit in a `wrong' potential, but as long as approximate spherical symmetry is preserved, their interaction rate with the BHs should not be significantly influenced by this (and during a close encounter, the background stellar potential is of course unimportant).

Assuming additionally that the interval is short enough so that the forces on the BHs due to the background stars do not vary significantly, we perform scattering experiments in each interval to find $\Delta E_\star$ and $\Delta L_\star$. Using the simple algebra described below, we find the average forces on the BHs over this interval. We also assume a purely planar motion of the BHs. Thus, unless stated otherwise, by `angular momentum' and `torque' we mean only the $z$ component of these vectors.

\subsection{Conservation laws}
The basic conservation equations are
\begin{align}
\Delta E_{{\rm BH}} &= -\Delta E_{\star}=\int_{{\bf s}_1(t_{0})}^{{\bf s}_1(t_1)}{\bf F}_1\cdot{\bf ds}_1 + \int_{{\bf s}_2(t_{0})}^{{\bf s}_2(t_1)}{\bf F}_2\cdot{\bf ds}_2,\label{eq:cons-E}\\
\Delta L_{{\rm BH}} &= -\Delta L_{\star}=\int_{t_{0}}^{t_1}\tau_1{\rm d}t + \int_{t_{0}}^{t_1}\tau_2{\rm d}t,\label{eq:cons-L}
\end{align}
where ${\bf F}_i$ and $\tau_i\equiv\left({\bf r}_i\times{\bf F}_i\right)\cdot\hat{{\bf z}}$ are the force and torque, respectively, exerted on BH number $i$ by the stellar population. Let us write ${\bf F}_i$ in the following form:
\begin{equation}
{\bf F}_i=-f_i\hat{{\bf v}}_i+\tilde{f}_i\hat{{\bf u}}_i,
\end{equation}
where $\hat{{\bf v}}_i$ and $\hat{{\bf u}}_i$ are the unit vectors parallel and perpendicular to the velocity of the BH. Note that $f$ is `drag like' and directed opposite to the velocity vector. If an object moves through a uniform background, symmetry dictates that the mean force would be antiparallel to the velocity vector. However in realistic environments there must be also a force due to the inhomogeneities of the background, so a perpendicular force component is required. In Cartesian coordinates:
\begin{align}
\hat{{\bf v}}_i &= \frac{v_{ix}}{v_i}\hat{{\bf x}}+\frac{v_{iy}}{v_i}\hat{{\bf y}},\\
\hat{{\bf u}}_i &= \frac{v_{iy}}{v_i}\hat{{\bf x}}-\frac{v_{ix}}{v_i}\hat{{\bf y}}.
\end{align}
To simplify equation (\ref{eq:cons-E}) we write ${\bf ds}_i=\hat{{\bf v}}_i{\rm d}s_i$ and thus
\begin{align}
\int_{{\bf s}_i(t_{0})}^{{\bf s}_i(t_1)}{\bf F}_i\cdot{\bf ds}_i &= 
\int_{{\bf s}_i(t_{0})}^{{\bf s}_i(t_1)}\left(-f_i\hat{{\bf v}}_i+\tilde{f}_i\hat{{\bf u}}_i\right)\cdot\hat{{\bf v}}_i{\rm d}s_i\notag\\
{} &=-f_i\int_{{\bf s}_i(t_{0})}^{{\bf s}_i(t_1)}{\rm d}s_i\equiv-f_i\mathcal{S}_i.\label{eq:def-S}
\end{align}
The perpendicular force component $\tilde{f}$ disappears due to the dot product, while $f$ is assumed constant along the path and can be taken out of the integral, which defines $\mathcal{S}$ (which is simply the path's length). Finally, equation (\ref{eq:cons-E}) for the energy becomes
\begin{equation}
\Delta E_{\star}=f_1\mathcal{S}_1+f_2\mathcal{S}_2.\label{eq:forces-E}
\end{equation}

To simplify equation (\ref{eq:cons-L}) we write $\tau_1$ in Cartesian coordinates:
\begin{equation}
\tau_i = -\frac{f_i}{v_i}(x_iv_{iy}-y_iv_{ix}) - \frac{\tilde{f}_i}{v_i}(x_iv_{ix}+y_iv_{iy}),
\end{equation}
and thus
\begin{align}
\int_{t_{0}}^{t_1}\tau_i{\rm d}t &= -f_i\int_{t_{0}}^{t_1}\frac{x_iv_{iy}-y_iv_{ix}}{v_i}{\rm d}t-\tilde{f_i}\int_{t_{0}}^{t_1}\frac{x_iv_{ix}+y_iv_{iy}}{v_i}{\rm d}t \notag \\
{} &\equiv -f_i\mathcal{P}_i-\tilde{f_i}\mathcal{Q}_i.
\end{align}
The above integrals define $\mathcal{P}$ and $\mathcal{Q}$. Equation (\ref{eq:cons-L}) for the angular momentum becomes
\begin{equation}
\Delta L_{\star}=f_1\mathcal{P}_1+\tilde{f_1}\mathcal{Q}_1+f_2\mathcal{P}_2+\tilde{f_2}\mathcal{Q}_2.\label{eq:forces-L}
\end{equation}

If we evolve the BHs between $t_0$ and $t_1$ under their mutual gravity alone, energy (and angular momentum) will be conserved along the produced orbital segment: $\Delta E_{\rm BH}=0$. Stars scattered on this orbital segment will not their conserve energy: $\Delta E_\star\neq 0$; so energy is also globally not conserved between $t_0$ and $t_1$. By solving equations (\ref{eq:forces-E}) and (\ref{eq:forces-L}) for the (non conserving) forces, the BHs can be evolved again in this time interval, under the additional forces, producing an orbital segment for which $\Delta E_{\rm BH}=-\Delta E_\star$. The revised orbital segment is only slightly different from the original, since the additional forces are much smaller than the forces exerted by the other BH and the background potential. Stars scattered on the revised orbit will have a slightly different $\Delta E_\star$. This process of alternatingly evolving the BHs and the scattering of stars can be repeated until converges is achieved.

Since $\Delta E_\star$ and $\Delta L_\star$ are directly obtained from the scattering experiments and the six integrals (calligraphic letters) are calculated from the orbital segments, equations (\ref{eq:forces-E}) and (\ref{eq:forces-L}) are a linear system of two equations with four variables: $f_1$, $f_2$, $\tilde{f}_1$ and $\tilde{f}_2$. It is worth noting that without the perpendicular force, which is expected to be negligible in the standard dynamical friction formalism, it is impossible to conserve energy and angular momentum simultaneously. In particular cases, additional constraints give an exact solution as explained in the following section.

\subsection{Finding the forces}\label{sec:forces}
\subsubsection{Symmetric motion}
In this case, the masses are equal and the initial conditions are symmetric with respect to the centre of the system. The motion of one BH mirrors that of its companion, so that the path integrals are equal for the two BHs (the index is therefore dropped) and the forces acting on them must also be equal due to symmetry. The solution is
\begin{align}
f &= \frac{1}{2\mathcal{S}}\Delta E_\star,\label{eq:force-sym-par}\\
\tilde{f} &= \frac{1}{2\mathcal{SQ}}(\Delta L_\star \mathcal{S}-\Delta E_\star \mathcal{P}).\label{eq:force-sym-per}
\end{align}
There is no solution for $\mathcal{Q}=0$ (but $\mathcal{S}\neq 0$ is guaranteed by equation \ref{eq:def-S}). In points where the BHs' velocity is purely tangential, $\tilde f$ is parallel to the radius vector and thus exerts no torque (and work is never done by $\tilde f$). At these points, $\tilde f$ is free but $f_i$ is overdetermined (must change both $E_{\rm BH}$ and $L_{\rm BH}$ by the specified amounts). Segments which are symmetric about such points have $\mathcal{Q}=0$.

\subsubsection{High mass ratio}
In this case one mass is much larger that the other, and is assumed to sit motionless at the centre of the system. The forces $f$ and $\tilde{f}$ act on the secondary BH only. The integrals $\mathcal{S}$, $\mathcal{P}$ and $\mathcal{Q}$ are also calculated for the secondary only. The solution for $\mathcal{Q}\neq 0$ is
\begin{align}
f &= \frac{1}{\mathcal{S}}\Delta E_\star,\\
\tilde{f} &= \frac{1}{\mathcal{SQ}}(\Delta L_\star \mathcal{S}-\Delta E_\star \mathcal{P}).
\end{align}

\subsubsection{General solution}
The results for the limiting cases described above motivate us to look for solutions of the form
\begin{align}
f_1 & =\frac{\alpha}{\mathcal{S}_1}\Delta E_{\star}, & \tilde{f}_1 & =\frac{\alpha}{\mathcal{S}_1\mathcal{Q}_1}\left(\Delta L_{\star}\mathcal{S}_1-\Delta E_{\star}\mathcal{P}_1\right),\\
f_2 & =\frac{1-\alpha}{\mathcal{S}_2}\Delta E_{\star}, & \tilde{f}_2 & =\frac{1-\alpha}{\mathcal{S}_2\mathcal{Q}_2}\left(\Delta L_{\star}\mathcal{S}_2-\Delta E_{\star}\mathcal{P}_2\right).
\end{align}
There is a mathematical solution for every $\alpha$, but we know from the limiting cases that $0\leq\alpha\leq\frac{1}{2}$. The value of $\alpha$ is a function of the mass ratio $q$, but it may also
be dependent on other factors, such as the local stellar densities at the instantaneous position of either BH. In the case of a high mass ratio, one BH is almost stationary; therefore $\alpha$ must approach zero faster or at least as fast as $\mathcal{S}_{1}$. The force $f_{1}$ does not have to approach zero, but the acceleration $f_{1}/M_{1}$ does.

\section{Model}\label{sec:model}
\subsection{Units}\label{sec:units}
For reasons of consistency with our previous work \citep{meiron10}, we use a unit system in which mass is measured in units of the primary BH's mass, velocity is measured in units of $4\sigma$ and $G=1$ (where $\sigma$ and $G$ are the stellar velocity dispersion and the gravitational constant, respectively). The hard binary separation is defined as
\begin{equation}
a_{\rm h} \equiv \frac{q}{1+q} \frac{GM_\bullet}{4\sigma^2},
\end{equation}
where $q \leq 1$ is the mass ratio of the secondary and primary BHs and $M_\bullet$ is the mass of the primary. In our units, the hard binary separation of an equal mass binary ($q=1$) is 2.

The base units are therefore scalable by  $M_\bullet$ and $\sigma$. Only one parameter is required if we also use the $M$--$\sigma$ relation (e.g. \citealt{m-sigma}). The units of length, time and velocity units and their scaling, using $M_\bullet$ and $\sigma$, are
\begin{align}
[{\rm L}] &= {\textstyle\frac{1}{16}} GM_\bullet\sigma^{-2} = 0.77\ M_8^{0.53} ~{\rm pc},\\
[{\rm T}] &= {\textstyle\frac{1}{64}} GM_\bullet\sigma^{-3} = 1\,000\ M_8^{0.29} ~{\rm yr},\label{eq:units-T}\\
[{\rm V}] &= 4\sigma = 750\ M_8^{0.24} ~{\rm km~s^{-1}},
\end{align}
where $M_8$ is the physical mass of the primary BH in units of $10^8 ~{\rm M}_\odot$. Note that in all simulations we used a stellar velocity dispersion of 0.25 velocity units.

\subsection{Bulge Properties}\label{sec:bulge}
In all our simulations, stars are initially distributed in a singular isothermal sphere and follow a Maxwell--Boltzmann distribution with 1D velocity dispersion $\sigma$. To avoid the non-physical divergence of the potential, we assume a core structure:
\begin{equation}
\rho(r)=
\begin{cases}
 \rho_0 & r\leq h\\
 \rho_0\left(\frac{h}{r}\right)^{2} & r>h
\end{cases}\label{eq:isothermal-density}
\end{equation}
where $h$ is an arbitrary break radius set to 1 and $\rho_0={\sigma^2}/{2\pi G h^2}$. The expression for the gravitational potential (or the bulge potential) derived from the above density is
\begin{equation}
\Phi_{\rm bulge}(r)=
\begin{cases}
 \frac{\sigma^2}{3Gh^2}r^2 & r\leq h\\
 \frac{\sigma^{2}}{G}\left[\frac{4h}{3r}+2\ln\left(\frac{r}{h}\right)-1\right] & r>h
\end{cases}\label{eq:bulge-potential}
\end{equation}
\citet{koopmans09} found that massive elliptical galaxies, within their effective radii, are well approximated by a power law ellipsoid with an index of $-2$. \citet{genzel03} also found that the 
density of the nuclear star cluster of the Milky way can be described by a broken power law with index of $-2.0 \pm 0.1$ down to 0.38 pc. This is however not appropriate within the BH sphere of influence, if dynamically relaxed, where the equilibrium distribution (radius independent mass and energy flow) is the famous \citet{bw76} cusp of $\rho \propto r^{-7/4}$ (where only one mass species is present, cf. \citealt{bw77}, \citealt{hopman-alexander06}). Unrelaxed clusters around adiabatically growing BHs are expected to have shallower slopes \citep{young80} or steeper slopes in the case of rotating systems and non-isothermal clusters \citep{lee89}.

In the case of a minor merger (equivalent to $q=0.1$), the structure of the more massive galaxy does not change significantly and equation (\ref{eq:isothermal-density}) likely represents the stellar environment seen by the secondary BH after its parent galaxy is absorbed. This picture is somewhat na\"ive in the case of a major merger ($q=1$), but violent relaxation \citep{lynden-bell67} due to the rapidly varying potential in a newly merged galaxy causes widening of the stellar energy distribution and is analogous to relaxation by collisions in a gas. This process tends to drive galaxies towards a universal steady state \citep{syer98}. A nice demonstration of this appears in the $N$-body merger simulations of \citet{mm01}, who find an $r^{-2}$ density profile at the time the equal mass binary becomes hard, which extends down to the scale of the binary separation.

Throughout this work we assume that the BHs are `naked', that is, do not carry clusters of bound stars. In the equal mass case, the stellar mass bound to a single BH once the binary becomes hard does not exceed 10 per cent of its mass (assuming that the cluster also has a power law density profile with index of -2 and normalization based on the $M$--$\sigma$ relation). This will most likely not affect the late time evolution, but will surely affect the early inspiral phase, and the exact inspiral time, but will probably not have a major effect.

\section{Algorithm}\label{sec:algorithm}
\subsection{Equal masses}\label{sec:alg-equal}
Below we provide a technical description of the application of the technique described in Section \ref{sec:math} for an equal mass binary simulation. The slightly different procedure for the 10:1 mass ratio case is discussed in the following Section.

First, a realization of a singular isothermal sphere is produced up to a cutoff radius of $\Rmax$, with two equal mass BHs placed on the $x$-axis at $x=\pm R_0$; at this initial distance the BHs are still unbound to each other and their inspiraling orbits are governed by the bulge's gravity and dynamical friction. The simulation duration is divided into (equal) intervals $\Delta t$, after each the stellar force acting on the BHs is updated. The actual time {\it steps} of the ODE solver are smaller if necessary, so close encounters can be resolved. Within each interval we do the following:
\begin{enumerate}
\item Symmetrically advance\footnote{The word `advance' in this context means solve the equation of motion by means of Runge--Kutta method of order five with adaptive step size control (e.g. \citealt{nr}).} the two BHs from $t_i$ to $t_{i+1}$, under their mutual gravity, and the effective forces exerted by the stars. The calculated antiparallel $f$ and perpendicular $\tilde{f}$ embody the stars' pull on the BHs, so the bulge potential (equation \ref{eq:bulge-potential}) should not be considered additionally. This gives a short orbital segment which is stored in memory.
\item Advance each star from $t_i$ to $t_{i+1}$, under the influence of the two BHs and the bulge potential. The motion of the BHs in this interval is already set from the previous stage, so this is essentially a short scattering experiment. This is the most computationally demanding stage of each iteration, but easily parallelized.
\item Sum up the energies and angular momenta of all stars and subtract the values from the previous iteration to obtain $\Delta E_\star$ and $\Delta L_\star$.
\item Use the BHs' path to obtain $\mathcal{S}$, $\mathcal{P}$ and $\mathcal{Q}$.
\item Calculate $f$ and $\tilde{f}$ from equations (\ref{eq:force-sym-par}) and (\ref{eq:force-sym-per}), to be used in the next $\Delta t$ interval.
\end{enumerate}

Unless otherwise indicated, our simulations end at $t=10\,000$ time units (or $10^7$ years for a $10^8~{\rm M}_\odot$ primary); this is equivalent to $\sim 1\,000$ revolutions after the binary becomes hard in most equal mass simulations. Integration of a single star is terminated prematurely in three cases: if it reaches a distance of $\Rmax+10$ length units from the centre of the system, reaches $r_{\rm tidal}=10^{-3}$ from either BH or takes more than 35 integration time steps to complete the interval of $\Delta t=0.1$ (see Section \ref{sec:qa}). In the first case the star is considered to have escaped the system (or {\it diverged}); the extra 10 length units beyond $\Rmax$ are an arbitrary `padding' required for technical reasons. The second case represents tidal disruption of the star (the orbit is then said to have {\it crashed}). The true tidal disruption radius is up to two orders of magnitude smaller than our $r_{\rm tidal}$, but this choice has a negligible effect on the BHs because the rate of crashing stars is negligible compared to the rate of diverging stars. The choice of $\Rmax$ is rather arbitrary, and is chosen to be large enough to minimize its effect on the results, as discussed in Section \ref{sec:Rmax}.

Notably, the forces on the BHs are one interval retarded. Thus, the interval duration must be short enough in order not to break conservation of energy and angular momentum. In all simulations we chose to use equal intervals of $\Delta t=0.1$, and since simulations done with this choice both ran reasonably quickly and performed well in terms of conservation, we did not thoroughly investigate changing $\Delta t$. Certainly making the interval length longer or adaptive can significantly speed up the simulations.

The code outlined above is very simple in terms of decision making; the bulk of CPU effort is made to individually advance stars (scattering experiments) with only a small non-parallelizable overhead between intervals. A typical simulation with a large realization ($N=5\times 10^6$) would run for $\sim$ few days on a medium strength personal desktop computer.

\subsection{Mass ratio}\label{sec:alg-ratio}
As noted in Section \ref{sec:forces}, the force acting on the BHs can be uniquely found from the energy and angular momentum differences only in the cases of equal masses and high mass ratios (where it is possible to assume that the primary BH is fixed at the centre). However, when the velocity is perpendicular to the radius vector (i.e. at pericentre or apocentre), the values of $\mathcal{Q}_i$ approach zero, and thus $\tilde{f}_i$ are left out of the coupled equations (\ref{eq:forces-L}) and (\ref{eq:forces-E}), and cannot be solved for. At the same time, $f_i$ are overdetermined as they have to compensate for both energy and angular momentum changes. Even in intervals that contain an apsis, there is usually a solution unless the orbital segment happens to be symmetric about the apsis. Since only a few segments are affected, the effect on the BBH orbits is not significant.

In the case of equal masses, no special treatment of these segments was necessary; but in the case of very high mass ratios (i.e. $q \ll 1$) the finite numerical fluctuations are larger due to the fact that the secondary BH is less massive. In these cases the total angular momentum would discontinuously drop at an apsis, and the otherwise smooth path would suddenly break at this segment. We compromised on simulating a 10:1 binary, and utilized two techniques to improve accuracy. First, each step was performed twice: after calculating the frictional force, the stars and BH were reset to their original positions, and advanced again with the newly calculated force on the BH. Second, we attempted to compensate for the accumulated error: instead of calculating the force components using purely $\Delta E_\star$ and $\Delta L_\star$ of the last $\Delta t$ interval, we added their respective accumulated errors (the correction term had weight of 0.1 per cent). Those adjustments dramatically improved the accuracy of the 10:1 simulations, but unfortunately we did not yet overcome all the technical problems associated with reliably  simulating a higher mass ratio inspiral from $\sim 50$ pc down to stalling separation.

\subsection{Quality Assurance}\label{sec:qa}
If the routine that advances the stars has a bug or is just not accurate enough, the values of $\Delta E_\star$ and $\Delta L_\star$ obtained in each interval will be faulty; but since the frictional force is calculated in such a way that would compensate for any change of $E$ and $L$ in the stars, the bug might remain undetected. Thus, an important validity test of $E$ and $L$ conservation in the code is inherently not available.

We can, however, test the ODE solver for a similar problem, and infer that our stellar orbits are at least {\it well behaved} in the original problem. Assume a BBH with a circular orbit of constant radius $R$; the BBH orbit is now decoupled from the stars and each stellar orbit can be integrated separately. This is the restricted 3-body problem plus a spherically symmetric potential; as in \citet{meiron10}, the energy in the rest frame of the BHs (the frame which rotates with the same angular frequency) is the only conserved quantity. The Jacobi integral is a constant of motion related to the rest frame energy by $C_J=-2E$.

We performed a number of tests where the BBH was in a circular orbit with a radius in the range of $1 \leq R \leq 40$ length units. The stellar model was the same as described in Section \ref{sec:bulge} and the orbits were evolved for $t=10\,000$ time units. It was found that $C_J$ is well conserved for the great majority of stars: only 0.2 to 1.6 per cent of orbits were cast off as {\it rogue} orbits, exceeding 35 steps per $\Delta t=0.1$ interval; the rest had an average $|\Delta C_J|$ of (4 to 11)$\times10^{-5}$ energy units ($C_J$ is typically of order unity). When applying a short softening length of 0.04 (corresponding to \citealt{qh97}), 0.2 to 1.7 per cent of orbits were eliminated by the same criterion and the rest had an average $|\Delta C_J|$ of (5 to 17)$\times10^{-5}$ energy units. Thus, we did not apply softening in the actual simulations. This type of number of steps filter against rogue orbits was found to work better than putting a lower limit on the allowed step size.

Accuracy may be improved simply by integrating close encounters with smaller error tolerance, or by employing more elaborate techniques such as regularization. The small number of stars which are lost due to the ODE solver's inability to handle them does not have a significant systematic effect on the BBH orbital evolution.

\subsection{Conservation of $E$ and $L$}
\begin{figure*}
\begin{center}
\includegraphics{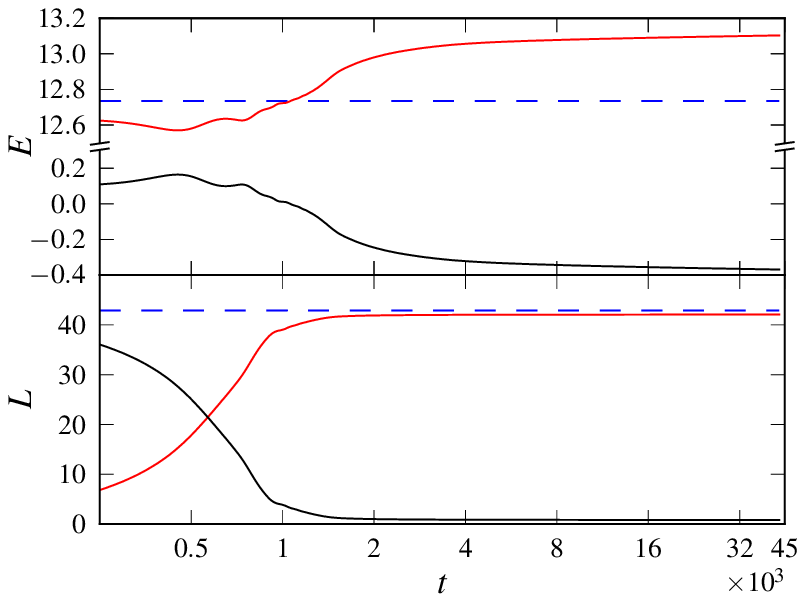}
\includegraphics{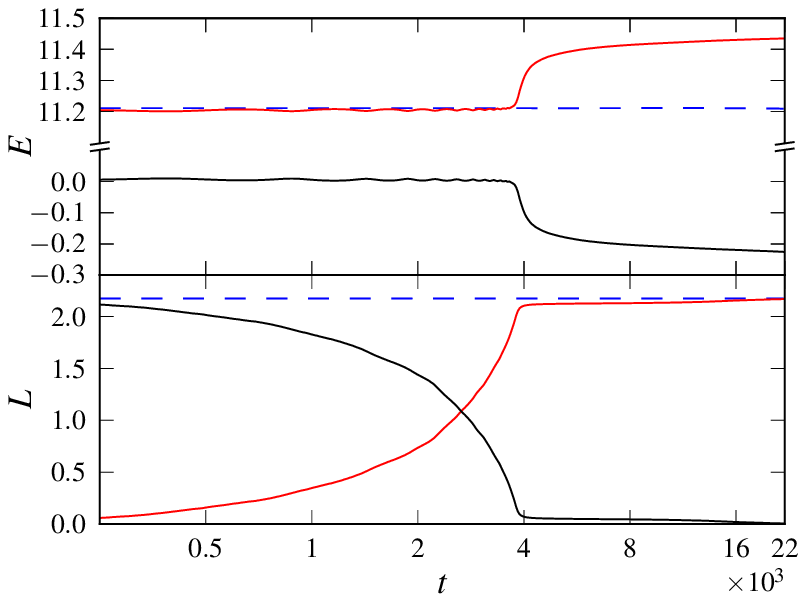}
\end{center}
\caption{Conservation of energy and angular momentum in the case of equal masses on a circular orbit (left panel; same simulation as shown in Fig. \ref{fig:long}), and the case of 10:1 mass ratio with initial eccentricity of 0.2 (right panel; same simulation as shown in Fig. \ref{fig:e-evolution}). The dashed blue lines are the system's total of the conserved quantities per unit mass; the solid red line represents the stellar population's part and the solid black line represents the BHs' part. The BH-star interaction term is in the stellar part, so the BHs' energy in this figure is only the sum of their kinetic energies and mutual potential energy.}
\label{fig:conservation}
\end{figure*}

As noted previously, total energy and angular momentum conservation is not a built-in requirement of the method, but rather indicates a successful transfer of energy and angular from the BBH to the stellar population. In the upper panels of Figure \ref{fig:conservation} we show energy of the BBH (solid black line), the stellar population (solid red line) and their sum (dashed blue line); the lower panels is the same but for the angular momenta. The left panel is for an equal mass simulation while the right panel is for a 10:1 mass ratio simulation.

The quality of energy conservation is attested by the absolute difference of total energy between the beginning and the end of the simulation; it is also useful to look at the fluctuations in total energy, a very crude estimate of which is the amplitude of the largest fluctuation (this is not a very good measure since fluctuation amplitude can be high in some parts of the orbit mild in others). These two quantities have dimensions of energy, and it is most sensible to normalize them with respect to the absolute difference in the energy of stars (or the BHs). Thus, for a simulation ending at $t=T$:
\begin{align}
\epsilon_1^E(T) &= [E_{\rm tot}(T) - E_{\rm tot}(0)] / |E_\star(T) - E_\star(0)|,\\
\epsilon_2^E(T) &= {\rm max}(|E_{\rm tot}(t) - \langle E_{\rm tot} \rangle|) / |E_\star(T) - E_\star(0)|.
\end{align}
We similarly define $\epsilon_1^L$ and $\epsilon_2^L$ for the angular momentum. All these $\epsilon$ parameters must be very small. If the quantity that is supposed to be conserved has some trend, then usually $\epsilon_1=\epsilon_2$, otherwise it only fluctuates around its mean value and $\epsilon_1<\epsilon_2$.

For an equal mass simulation with $N=5\times 10^6$ (discussed in detail in Section \ref{sec:N}) case at $T=43\,200$, we get $\epsilon_1^E, \epsilon_1^L < 10^{-6}$ and $\epsilon_2^E = \epsilon_2^L = 3\times 10^{-4}$. For a 10:1 mass ratio simulation (discussed in detail in Section \ref{sec:e-growth}) at $T = 21\,800$, we get $\epsilon_1^L=4\times 10^{-6}$ and $\epsilon_2^L=6\times 10^{-5}$, with a decreasing trend in total energy: $\epsilon_1^E = \epsilon_2^E=0.008$. The trend in total energy begins at $t\sim 14\,000$ and is probably due to the fact that there is an accumulated inaccuracy in the solution of the BHs' equations when they are so close together at pericentre. At $t = 14\,000$ we get $\epsilon_1^E = 8\times 10^{-5}$ and $\epsilon_2^E = 4\times 10^{-4}$ with values for the angular momentum similar to the end of the simulation. Thus, the numerical effect previously discussed is unrelated the eccentricity growth observed in this simulation.

\section{Results}\label{sec:results}
\begin{figure}
\includegraphics{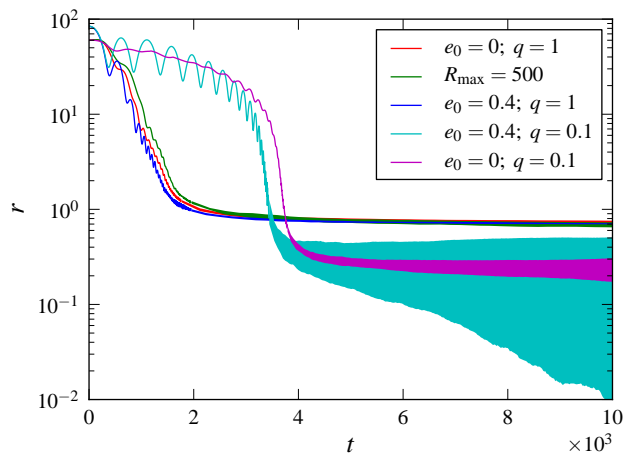}
\caption{A representative sample of simulation results showing the radius as a function of time. The red and blue lines are for an initially circular and eccentric equal mass binaries, respectively; magenta and cyan are the same for a mass ratio of 10:1; the green line is the same as the red line but with a larger cutoff of the stellar bulge. The thickness of the line is indicative of eccentricity. While the semi-major axis stalls in all simulations, eccentricity does not reach a steady state in the unequal mass cases.}
\label{fig:examples}
\end{figure}

We performed a total of 58 simulations of an equal mass binary and a 10:1 mass ratio binary, varying $N$, $R_0$, $\Rmax$ and the initial eccentricity $e_0$. In Figure \ref{fig:examples} we shows $r(t)$, for selected simulations. In the 10:1 simulations, the primary BH is fixed at the centre, so $r$ is the BBH separation; in the equal mass simulations $r$ is half the separation. As can be seen in the figure, the semi-major axis stalls in all simulations but eccentricity (indicated by the thickness of the lines) does not reach a steady state in the unequal mass cases. These results are discussed in more detail below.

\subsection{Number of Particles}\label{sec:N}
Here we show two things: how the results converge with increasing number of stars $N$, and how the results depend on the specific realization of the stellar distribution. We present seven pairs of simulations with $N$ between $50\,000$ and $350\,000$; for each $N$ the two simulations have a different random seed, so that the stars have different initial positions and velocities, but are drawn from the same distribution. All simulations are of an equal mass binary starting at $R_0=60$ with the local circular velocity; the cutoff radius of the stellar sphere is $R_{\rm max}=200$.

\begin{figure}
\includegraphics{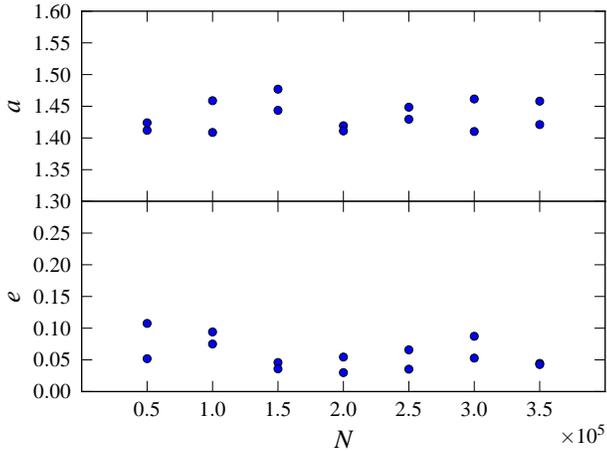}
\caption{The semi-major axes $a$ and eccentricities $e$ at $t=10\,000$ as functions of the number of particles $N$, for equal mass and initially circular binaries. For each $N$, the two circles represent two different realizations of the same stellar population. The effect of changing $N$ in the range tested in this set is comparable in magnitude to that of changing the realization, and no trend is seen. The simple averages for the entire set are $a = 1.43 \pm 0.02$, corresponding to a stalling separation of $\sim 1.1$ pc for two $10^8~{\rm M_\odot}$ BHs, and $e = 0.06 \pm 0.02$ which indicates nearly circular orbits. An additional run with $N = 5\times 10^6$ is not shown here (see text).}
\label{fig:N}
\end{figure}

In Figure \ref{fig:N} we show the `final state' (i.e. at $t=10\,000$) semi-major axes $a$ (equivalent to separation, in the equal mass case), and eccentricities $e$ as functions of $N$; these two numbers are the best way to appreciate differences between similar simulations. From this small sample, it is apparent that the effect of changing $N$ in this range is comparable in magnitude to that of changing the realization. In this set of simulations, the range of semi-major axis values is 0.07 length units or $\sim 5$ per cent of the sample's average $a$; the eccentricities are small and in the range $0.03 < e < 0.1$. Thus, increasing $N$ beyond $100\,000$ (within $R_{\rm max}=200$) is unnecessary for this level of accuracy and following tests are made using this number.

The average of $a$ in these simulations is 1.43 (in physical units, for $10^8 ~{\rm M}_\odot$ BHs, this is equivalent to 1.1 pc); this is approximately 30 per cent below the hard binary separation of 2. \citet{merritt06} suggested the following formula for the stalling separation:
\begin{equation}
\frac{a_{\rm stall}}{r_h^\prime} = 0.2 \frac{q}{(1+q)^2}\label{eq:stall},
\end{equation}
where ${r_h^\prime}$ is the radius containing a mass in stars equal to twice the combined mass, or $M({r_h^\prime})=2M_\bullet(1+q)$, at the time of stalling. We measured the accumulated mass in the above simulations and got that in all of them ${r_h^\prime}=37.5$ with a very small spread. By substituting this information and $q=1$ into equation (\ref{eq:stall}), one gets $a_{\rm stall}=1.88$. It is important to note that in the \citet{merritt06} simulation there was no actual stalling of the BBH, and $a_{\rm stall}$ was estimated as the value of $a$ in which a clear change in the hardening rate took place; ${r_h^\prime}$ was determined at the time when this change occurred. In our simulations both ${r_h^\prime}$ and $a_{\rm stall}$ were determined at $t=10\,000$, which is well after the hardening rate has dropped. Also, equation (\ref{eq:stall}) was calibrated by \citeauthor{merritt06} using simulations with mass ratios $0.025 \leq q \leq 0.5$.

\begin{figure}
\includegraphics{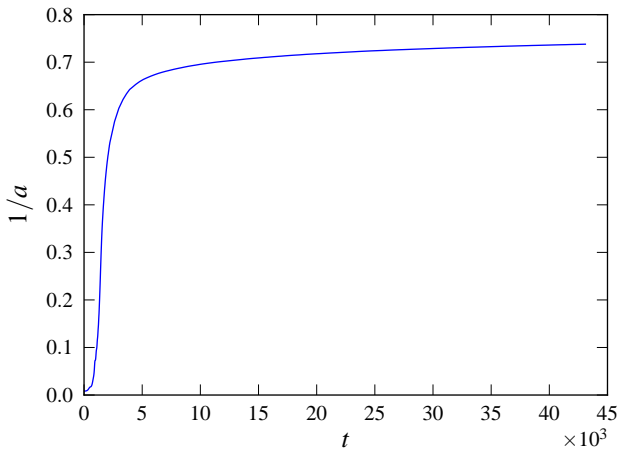}
\caption{The inverse semi-major axis of an equal mass binary simulation that was performed with $N = 5\times 10^6$ particles (compared with $N \leq 3.5\times 10^5$ for the simulations in Fig. \ref{fig:N}) and for a much longer duration. The value of $a$ at $t=40\,000$ is 1.35, which is 5.6 per cent lower that the value at $t=10\,000$ and only one per cent lower that the value at $t=30\,000$. Cf. figure 1 of \citet{berczik06}}
\label{fig:long}
\end{figure}

To test the convergence of the results, we also performed a single simulation with significantly more particles ($N = 5\times 10^6$) and longer duration ($t = 43\,200$); we show the inverse semi-major axis as a function of time in Figure \ref{fig:long}. As seen in the figure, there is still some slow evolution of the semi-major after $t=10\,000$. At $t=40\,000$ the value of $a$ is 1.35, which is 5.6 per cent lower than the value at $t=10\,000$ and only one per cent lower than the value at $t=30\,000$; the decay rate at the end of this simulation is $\dot{a}=10^{-6}$ velocity units, equivalent to $\sim 10^{-9}~{\rm pc\ yr^{-1}}$ for a $10^8~{\rm M}_\odot$ primary. In this specific run, eccentricity was especially low at $e<0.01$.

The orbits produced in the other simulations in this set (with $N\leq 3.5\times 10^5$) are very similar to the orbit shown in Fig. \ref{fig:long}, regardless of the number of particle (cf. \citealt{berczik06}, spherical and triaxial cases). Note that a large $N$ is required only to minimize the statistical fluctuations, our scheme is not subject to an artificial stellar relaxation mechanism, which requires a very large $N$ to overcome.

\subsection{\textbf{\emph{R}}${}_{\bf max}$ and \textbf{\emph{R}}${}_{\bf 0}$}\label{sec:Rmax}
Since we only simulate the spherical component and not a full merger, the initial and boundary conditions need be assumed: the initial distance of the BHs from the centre, $R_0$, and the cutoff radius of the stellar sphere, $\Rmax$. Here we present six trios of simulations with $\Rmax$ between 120 and 520, each three simulations are of an equal mass binary starting at $R_0=30$, $60$ and $90$ with the local circular velocity. The number of particles is $N = 500 \times \Rmax$, so as to keep the particle density profiles of equal normalizations (but different cutoff radii) in all simulations; this is due to the fact that the number of particles (or mass) grows linearly with radius in an isothermal sphere.

\begin{figure}
\includegraphics{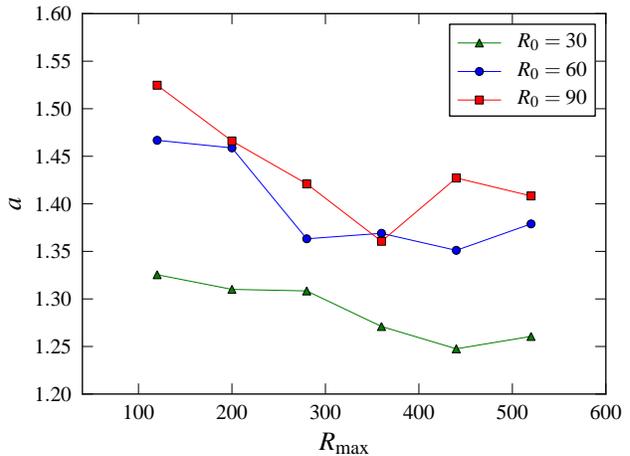}
\caption{The semi-major axes $a$ at $t=10\,000$ as a function of the cutoff radius $\Rmax$, for equal mass and initially circular binaries. For each of $\Rmax$, the tree data points represent BHs launched from different radii $R_0$. The increase of $a$ with $R_0$ seems to saturate, while the dependence on $\Rmax$ is very weak in the tested range. Eccentricity in this simulation was not correlated with either of the tested parameters, its simple average for the entire set is $e = 0.06 \pm 0.03$ which indicates nearly circular orbits.}
\label{fig:R}
\end{figure}

In Figure \ref{fig:R} we show the final state semi-major axes $a$ (stalling separation) as a function of $\Rmax$. From the results discussed in Section \ref{sec:N}, a characteristic error of $\sim 0.04$ units on the semi-major axis can be attributed to a specific realization and number-of-particles statistical fluctuation. The simulations with $R_0=30$ give systematically lower values for $a$; the values for $R_0=60$ do appear to be systematically lower than the $R_0=90$ simulations, but the two sets are within the errors of each other. There appears to be a trend of decreasing values of $a$ with increasing $\Rmax$ up to $\sim 300$, which is to be expected both because there is a larger supply of particles that can interact with the BHs, and because the potential well is deeper and stars that have already interacted with the BHs have a larger probability to fall back to the centre and interact again. Nevertheless, we see from this small sample that these effects are weak and comparable in magnitude to those discussed in the Section \ref{sec:N}. In this set of simulations, the eccentricities are also small and in the range $0.02 < e < 0.11$; no correlation of eccentricity was observed with either $R_0$ or $\Rmax$.

\subsection{Mass Ratios}\label{sec:mass-ratios}
As noted in section \ref{sec:alg-ratio}, the simulation of an unequal mass binary is somewhat different in nature. Thus, when studying the evolution of a 10:1 binary, we followed the analysis of Section \ref{sec:N} and performed a number of different tests with increasing number of particles and different realizations. Here we present four pairs of simulations with $N$ between $100\,000$ and $400\,000$; for each $N$ the two simulations have a different random seed, so that the stars have different initial positions and velocities, but are drawn from the same distribution. All simulations are of a binary with a 10:1 mass ratio starting at $R_0 = 60$ with the local circular velocity; the cutoff radius of the stellar sphere is $\Rmax = 200$.

\begin{figure}
\includegraphics{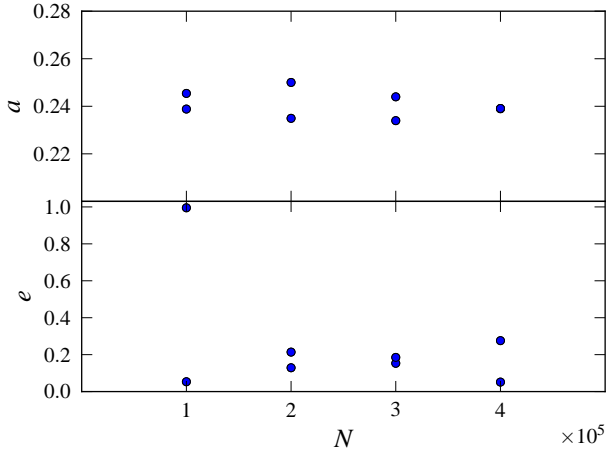}
\caption{Same as Fig. \ref{fig:N} but for a 10:1 binary. 
The semi-major axes are consistent within this sample, the simple average is $a = 0.241 \pm 0.005$, corresponding to a stalling separation of $\sim 0.19$ pc for for a $10^7~{\rm M_\odot}$ secondary around a $10^8~{\rm M_\odot}$ primary. Eccentricity varies greatly in this sample, see text.}
\label{fig:M}
\end{figure}

In Figure \ref{fig:M} we show the final state semi-major axes $a$ and eccentricities $e$ as functions of $N$. As in Section \ref{sec:N}, changing $N$ in this range produces no apparent effect on the stalling radius. Within this set of simulations, the semi-major axes are consistent; the sample's average $a$ is $0.241 \pm 0.005$. We measured the accumulated mass in the above simulations and got that in this case ${r_h^\prime}=16.8$ with a very small spread. By substituting this information and $q=0.1$ into equation (\ref{eq:stall}), one gets $a_{\rm stall}=0.28$. The caveats of using this equation were explained in Section \ref{sec:N}. If we recalibrate equation (\ref{eq:stall}) using our two values of $q$, the prefactor is lowered from 0.2 to approximately 0.16; it is even somewhat smaller considering the fact that the true value of $a_{\rm stall}$ can be $\sim 6$ per cent lower than its measured value at $t=10\,000$ (see Figure \ref{fig:long}).

In contrast with the equal mass case, the eccentricities do not reach a steady state value. In one of the $N=10^5$ simulations, the eccentricity increases very rapidly after the binary becomes hard. At $t=9\,400$ the pericentre distance reached $10^{-3}$ and the simulation is terminated. In the rest of the simulations, The eccentricity values range between 0.05 and 0.28 at $t=10\,000$, however, in all but one of these simulations eccentricity is slowly increasing. We calculate $T_{e\rightarrow 1}$, a very rough estimation for the time of gravitational wave regime, by fitting $e(t)$ with a linear function in the range $6\,000<t<10\,000$ and continue it to $e=1$. At this time span the rise in eccentricity is approximately linear, however this trend breaks at approximately $e=0.85$, so in fact $T_{e\rightarrow 1}$ can underestimate the time to the gravitational wave regime by some $\approx 20$ per cent. In Table \ref{tab:e} we show the results for $T_{e\rightarrow 1}$ for this simulation set and also for the initially eccentric runs. The values range between approximately $4\times10^4$ and $3\times10^5$ time units, or on the order of $10^8$ years for a $10^7~{\rm M_\odot}$ secondary around a $10^8~{\rm M_\odot}$ primary.

\begin{table}
\caption{A list of all 10:1 binary simulations. The columns from left to right are: the number of particles, initial eccentricity, eccentricity at $t=10\,000$ and the estimated timescale to reach $e=1$ (see text for definition).}
\label{tab:e}
\begin{tabular}{llll}
\hline
$N$ [$10^5$] & $e_0$ & $e_{\rm f}$ & $T_{e\rightarrow 1}$ \\
\hline
1 & 0   & 0.05     & $2.7\times 10^5$ \\
1 & 0   & $\sim 1$ & $\lesssim 10^4$  \\
2 & 0   & 0.21     & $4.8\times 10^4$ \\
2 & 0   & 0.13     & $2.2\times 10^5$ \\
3 & 0   & 0.15     & $1.4\times 10^5$ \\
3 & 0   & 0.18     & $5.9\times 10^4$ \\
4 & 0   & 0.05     & $4.4\times 10^7$ \\
4 & 0   & 0.28     & $3.5\times 10^4$ \\
2 & 0.1 & 0.42     & $2.0\times 10^4$ \\
2 & 0.1 & 0.33     & $2.5\times 10^4$ \\
2 & 0.2 & 0.47     & $2.1\times 10^4$ \\
2 & 0.2 & 0.59     & $1.7\times 10^4$ \\
50& 0.2 & 0.52     & $1.8\times 10^4$ \\
2 & 0.3 & 0.21     & $9.2\times 10^4$ \\
2 & 0.3 & $\sim 1$ & $\lesssim 10^4$  \\
2 & 0.4 & 0.96     & $\approx 10^4$ \\
2 & 0.4 & $\sim 1$ & $\lesssim 10^4$ \\
\hline
\end{tabular}
\end{table}

\subsection{Eccentricity growth}\label{sec:e-growth}
Here we tested how the initial eccentricity affects the results. We performed simulations with four different initial eccentricities between 0.1 and 0.4 with semi-major axes as in the circular simulations of Sections \ref{sec:N} and \ref{sec:mass-ratios}. The initial eccentricity $e_0$ is that of a BBH with the same initial conditions and that moves in the initial potential (equation \ref{eq:bulge-potential}) with no friction; the orbit is not really an ellipse since the potential in not Keplerian, so $e_0$ corresponds to the mean orbital eccentricity, defined as the difference between the maximum and minimum separations divided by the major axis. For each $e_0$, there are two simulations (with two different realizations) for an equal mass binary and two for a 10:1 binary. The number of particles is $N = 2\times 10^5$ with cutoff radius $\Rmax=200$; we performed a single simulation with significantly more particles ($N = 5\times 10^6$) for the $e_0=0.2$ case.

\begin{figure}
\includegraphics{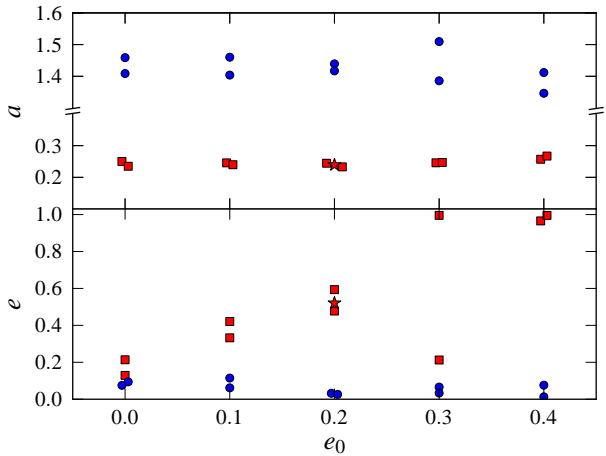}
\caption{The semi-major axes $a$ and eccentricities $e$ at $t=10\,000$ as functions of initial eccentricity $e_0$, for equal mass (blue circles) and 10:1 binaries (red squares). For each $e_0$, the two data points of each kind represent two different realizations of the same stellar population (all simulations have discrete values of $e_0$, but some overlapping data points were moved slightly to the left and right for graphical reasons). In the 10:1 case, the final eccentricities increase with $e_0$ while the semi-major axes are unaffected. The equal mass binaries seem to circularize independently of their initial eccentricities, as all results are very similar in both $a$ and $e$. The red stars represent a single 10:1 simulation with $N=5\times 10^6$; all others shown here have $N=2\times 10^5$.}
\label{fig:E0}
\end{figure}

In Figure \ref{fig:E0} we show the final state semi-major axes $a$ and eccentricities $e$ as functions of $e_0$. This figure also includes four data points with $e_0=0$ that have already been presented in Figs. \ref{fig:N} \& \ref{fig:M}. The stalling separation is independent of initial eccentricity in the tested range, the sample's average $a$ is $1.42 \pm 0.04$ for the equal mass simulations and $0.246 \pm 0.010$ for the unequal mass simulations. As in the other tests we performed for equal mass binaries, the final orbits are very much circular; the eccentricities are in the range $0.01 < e < 0.1$ despite the initially significant eccentricity. However, the final eccentricities in the 10:1 cases do appear to be generally correlated with $e_0$. As in Section \ref{sec:mass-ratios}, here too, eccentricities are still increasing when the simulations terminate at $t=10\,000$; the eccentricity timescales $T_{e\rightarrow 1}$ for the 10:1 simulations are shown in Table \ref{tab:e}, which shows that the eccentricity growth rate is related to $e_0$.

\citet{sesana10} found that eccentricity growth is generally mild for equal mass binaries with very small initial eccentricities, but also that initial eccentricity $e_0 > 0.3$ leads to very high peak eccentricity almost regardless of the other system parameters. Although all our equal mass simulations end up in nearly circular orbit, this is not inconsistent with \citeauthor{sesana10}: while in the latter work the BHs are launched from within their radius of influence, in our simulations the BHs are launched from much further out, in the dynamical friction regime. In our simulations, an equal mass binary becomes less eccentric as it inspirals from $r>60$. For one of the $e_0=0.4$ simulations, the rapid eccentricity decrease ceased at about $r=1.4$ (approximately twice the stalling radius) where the value was $e=0.035$. At $r=4$ (equivalent to the initial radius of \citeauthor{sesana10}'s equal mass binaries) the eccentricity was 0.12; no significant eccentricity growth occurred in \citeauthor{sesana10}'s equal mass simulation with initial eccentricity of 0.1. While it has been shown by \citet*{dotti06} that BBHs lose memory of their initial eccentricity if they corotate with a massive gaseous disc, studies of eccentric orbits of hard binaries is motivated for the purely stellar dynamical case by the theory of linear response for dynamical friction \citep*{colpi99}. However, this theory is derived from a first order perturbative expansion and is not applicable when close encounters dominate the evolution, and the system is not well described by an analytical approximation.

It is important to note another major difference between our work and \citet{sesana10} which greatly affects the evolution of the binary orbital parameters, mostly the semi-major axis: while in our work the loss cone empties, \citeauthor{sesana10} implicitly assumes that the loss cone is always full at $r > r_{\rm inf}$. This leads to a very rapid decay of the binary separation and quick coalescence due to gravitational wave emission. Similarly, \citet{antonini12} calculated the eccentricity evolution in the case of a very small secondary BH that does not affect the stellar density profile. There, eccentricity grows because the orbit passes in and out of a flat core, where the star are fast and the drag force is much less efficient at pericentre than at apocentre. In our simulations, however, the much more massive secondary forms a cavity slightly larger than its apocentre, and eccentricity grows where the density is essentially zero.

\begin{figure}
\includegraphics{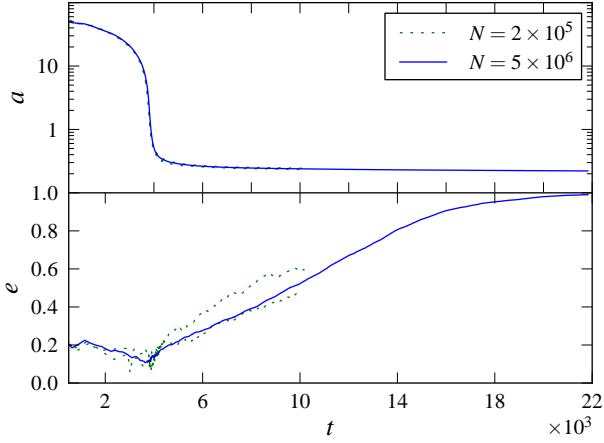}
\caption{The semi-major axes $a$ and eccentricities $e$ at as functions of time, for three 10:1 binary simulations with initial eccentricity of 0.2. The dotted green lines are two different realizations with $2\times 10^5$ particles while the solid blue line is a run with $5\times 10^6$ particles. The first two simulations are arbitrarily terminated at $t=10\,000$ while the last is stopped only when the eccentricity reaches 0.99, this time for a $10^8~{\rm M_\odot}$ primary is 22 Myr from the beginning of the simulation. For this mass scaling, the timescale for coalescence due to gravitational waves at the end of the simulation is just 1 Myr.}
\label{fig:e-evolution}
\end{figure}

In Figure \ref{fig:e-evolution} we show the evolution of the semi-major axis for all the runs with $e_0=0.2$ and 10:1 mass ratio. The dotted green lines represent the two realizations with $2\times 10^5$ particles while the solid blue line represents the larger $N=5\times 10^6$ realization. The first two simulations are arbitrarily terminated at $t=10\,000$ while the latter is stopped only when the eccentricity reaches 0.99. If we scale to physical units for a primary BH of $10^8~{\rm M_\odot}$, the end of the simulation is 22 Myr from its beginning. For this mass scaling, using the \citet{peters64} formula for orbital decay due to gravitational waves, the timescale for coalescence at the end of the simulation is just 1 Myr.

The rapid growth of the eccentricity while the semi-major axis remains fairly constant indicates a high value for the the dimensionless eccentricity growth, defined as:
\begin{equation}
K = \frac{{\rm d}e}{{\rm d}\ln (1/a)}.
\end{equation}
For single scattering of unbound stars from a fixed background, \citet{quinlan96} derived a maximal value of $K$ of about 0.3 for mass ratio of 16:1, consistent with previous scattering experiments (\citealt{roos81}; \citealt{mikkola92}). The value of $K$ for the simulation presented in Fig. \ref{fig:e-evolution} is at least an order of magnitude larger (we only roughly estimated the value from the results).
The difference is probably due to the very different nature of the orbits in the restricted 3-body problem versus the more realistic model used here. We refer an in-depth study of the physical mechanism behind the eccentricity growth to a subsequent work.

It is interesting to compare our results to those of \citet{iwasawa11}, who also got `runaway' growth of eccentricity while the semi-major axis stalled, but one must note the critical differences between the two studies. Most notably, \citeauthor{iwasawa11} used a mass ratio of 100:1 while we took only 10:1. Additionally, they used an initially very shallow central density profile, $\rho \propto r^{-3/4}$ while our bulge model (see Section \ref{sec:bulge}) was an isothermal sphere, $\rho \propto r^{-2}$. More importantly, the mass of their entire stellar population was just $1.25\times 10^9~{\rm M_\odot}$, which is 8 times less than the primary BH's mass and only 12.5 times more than the secondary's mass. By comparison, the total stellar mass in our model is 25 times the mass of the primary and 250 times the mass of the secondary. Thus, their entire simulation is deep within the primary's sphere of influence, where its gravity dominates, while our simulations started with the secondary well outside the primary's radius of influence.

Let us scale our work to theirs by setting $M_8 = 100$ in the scaling equations of Section \ref{sec:units}. \citeauthor{iwasawa11} start their {\tt A32k} simulations (with $N=32\,768$) at $R_0 = 20~{\rm pc}$ within which there are less than $2\,500$ particles, their BH stalls at $a = 3.9~{\rm pc}$. We initiate the secondary BH at $r=637~{\rm pc}$ and get stalling at 2.1 pc; there are initially $40\,000$ particles enclosed within $r = 20~{\rm pc}$ in our large $N=5\times 10^6$ simulation (marked with a star in Fig. \ref{fig:E0}). The secondary BH is $20\,000$ times more massive than a field star in our simulation, versus $2\,600$ in the \citeauthor{iwasawa11} simulation. Their stellar bulge model is very small compared to ours, with 90 per cent of their mass is within $R_0 = 190~{\rm pc}$, while our model is truncated at $\Rmax = 1\,800~{\rm pc}$.

The semi-major axes in the two studies evolve at very different rates: while the \citeauthor{iwasawa11} binary takes 19 million years (Myr) to sink from $a_0 = 20~{\rm pc}$ to 10 pc in their shallow cusp, our binary does the same journey in only 0.18 Myr. However, when applying equation (6) of \citet{iwasawa11} (derive from their numerical results) for our physical parameters, the timescale for significant eccentricity growth is extremely short at 0.78 Myr; the lifetime of the system from our simulation is approximately 85 Myr (scaled with $M_8 = 100$), which is two orders of magnitude longer. The difference might be due to the mass scaling assumed in their formula.

A final note about precession: in the unequal mass case, the BHs exhibit very small precession during the hard phase. In the large simulation of Fig. \ref{fig:e-evolution}, between $t_1=10\,000$ and $t_2=11\,000$ the semi-major axis precesses by $0.311^\circ$. During this period of 1 Myr (scaled with $M_8 = 1$), the average semi-major axis is 0.18 pc, and it drops by 0.002 pc; the average eccentricity is 0.559 and the growth is by 0.072. This precession can be produced, for example, by perturbing the Keplerian potential with a uniform density field of ~20,000 solar masses per parsec cubed. Even with our large number of particles, this density corresponds to only four particles enclosed in the sphere with radius equals to the apocentre. In the snapshot taken at $t_1$, there was one particle inside this region, and it was probably a transient since there are no stable orbits there except those tightly bound to one of the BHs. In principle, a small flux of particles to this region can produce the measured precession, but torques due to the anisotropy of the potential at larger distances are more likely to cause the precession. For comparison, general relativistic precession of the orbit (not reproduced in the simulation) would be $56^\circ~\rm{Myr}^{-1}$. We cannot yet say whether this precession compromises resonances that possibly induce the eccentricity growth, but will refer to this point in a future paper.

\section{Dynamical friction}\label{sec:df}

The orbital decay of a massive object within a galaxy down to its centre is well approximated by Chandrasekhar's dynamical friction formula \citep{chandra}. However, the assumption of a uniform background in his classical treatment does not hold in real galaxies. Thus, corrections to the Coulomb logarithm are necessary to account for the changing background with radius. \citet{just05} performed a detailed theoretical investigation of Chandrasekhar's formula in the presence of a density gradient and gave an improved analytical formula for the Coulomb logarithm. \citet{just11} took into account also self consistent velocity distribution functions, and made a comprehensive examination of the applicability of the new formula to sinking massive objects in a number of galaxy models, using high resolution $N$-body and particle-mesh codes. Their results suggest a delay in the orbital decay with respect to the standard formula, which quantitatively varies according to the studied case.

They notably give an explicit solution for the decay of a massive object moving on a `circular' orbit. Their formula (equation 25 in their paper) is very general and holds for an arbitrary power law density profile (it does not hold for very flat cores where fast moving stars contribute to most of the frictional drag; see \citealt{antonini12}). Here we bring their formula in our model's units and adjusted the parameters for an isothermal sphere; the radial evolution is given implicitly by:
\begin{equation}
\textstyle t = 52.918 \times \left\{{\rm Ei}\left[2\ln\left(\frac{1}{8}R_0\right)\right]-{\rm Ei}\left[2\ln\left(\frac{1}{8}r\right)\right]\right\}\label{eq:Just}
\end{equation}
The prefactor is an exactly calculable number. The special function ${\rm Ei}(x)$ is called the exponential integral, it has real values only for $x>0$. Thus, the domain of definition of equation (\ref{eq:Just}) is $r>8$. However, the assumption of a nearly circular orbit breaks well above $r=8$. The angle between the velocity vector and the tangent can be derived by finding the radial velocity from equation (\ref{eq:Just}):
\begin{equation}
\theta = 195.99^\circ \times \frac{1}{r}\ln\left({\textstyle\frac{1}{8}}r\right),\label{eq:spiral-angle}
\end{equation}
where this approximation hold only for small angles or large $r$.

It is generally difficult to simulate a full infall of a compact object into a galaxy centre due to the collisional nature of the interactions and the low number density of particles which can generally be obtained at the outskirts of galaxy models. \citet{just11} used the particle-mesh code {\sc Superbox} \citep{superbox}, which is collisionless and uses fixed time steps; This type of code, unlike direct $N$-body codes, allows a large particle number to be simulated in a relatively short time. Using the code we designed for BBHs, we can also study the early part of the inspiral, which is dominated by dynamical friction. Since our code resolves collisions between the BHs and stars, this study is complimentary to that of Just et al..

\begin{figure}
\includegraphics{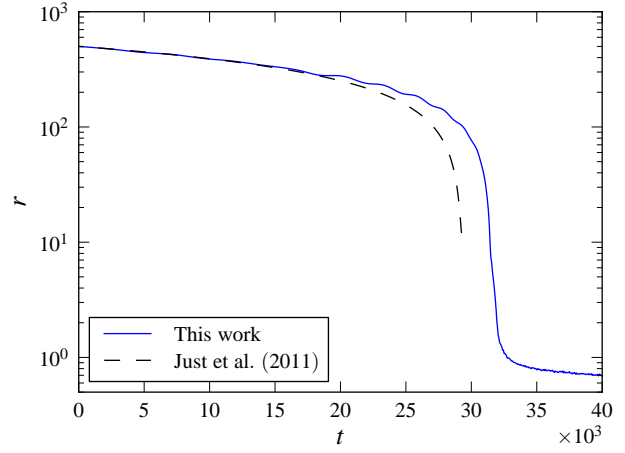}
\caption{This simulation of an equal mass binary initiated from $R_0=500$ (solid blue line) focuses on the dynamical friction regime. The theoretical curve (dashed black line) is the analytical formula of \citet{just11} given by equation (\ref{eq:Just}). The orbital decay is initially very well described by the formula, and the deviations at $r\lesssim 300$ are possibly due to inconsistency of the phase space distribution in the actual simulation versus the assumption of isothermal sphere used to derive the formula}
\label{fig:df}
\end{figure}

In Figure \ref{fig:df} we show a simulation of an equal mass binary initiated from $R_0=500$ (solid blue line) and the theoretical curve (dashed black line), equation (\ref{eq:Just}). The initial velocity is the local circular velocity, but the initial velocity vector is tilted by $1.62^\circ$; this angle is obtained be substituting $r=R_0$ in equation (\ref{eq:spiral-angle}). If the initial velocity is purely tangential, then the spiral becomes slightly `eccentric'. A second run with a different realization (not shown) gave very similar results, including the position of the wiggles. It should be pointed out that this simulation is just a proof of concept; we do not expect the stellar distribution to be spherically symmetric at the early stages of a major merger.

The orbital decay in our simulation is initially very well described by equation (\ref{eq:Just}); deviations become significant below $r \sim 300$, where mutual gravity of the two BHs is still negligible compared to the gravity of the bulge. These deviations are possibly due to the fact that the phase space distribution in the actual simulation (at the time and radius where the deviations occur) is no longer consistent with the assumption of isothermal sphere used to derive equation (\ref{eq:Just}), in particular the velocity distribution might not be described well by a Gaussian.

\section{Summary}\label{sec:summary}

Using a conservation-based scheme, we were able to follow the evolution of a BBH from a wide separation (enclosed stellar mass greater than the combined BH mass) down to sub-parsec scale. Our code resolves star-BH collisions and can run with $N>10^6$ stars on a desktop computer. We verified that our scheme yields convergent results which are independent of the number of particles, and the initial and boundary conditions. By performing scattering experiments on the inspiraling BBH, we will be able to extend \citet{meiron10} and calculate the signature of the inspiral on the background stellar phase space distribution as a function of projected position. This calculation improves on $N$-body simulation by reducing statistical fluctuations and having no spurious relaxation (and thus no loss cone refilling). We performed calculations for both an equal mass binary and a 10:1 mass ratio. Our calculations reveal:
\begin{enumerate}
\item The inspiral from a radius scale of tens of parsecs to the hard binary radius occurs on a time scale of a few million years for a $10^8 ~{\rm M}_\odot$ primary, with only a weak dependence of the timescale on the mass ($\propto M_8^{0.29}$, equation \ref{eq:units-T}).
\item The inspiral ends at a radius which is $\sim 30$ per cent smaller than the simple analytical estimate for the hard binary radius, and consistent with \citet{merritt06}.
\item An equal mass binary inspiral leads to a nearly circular final orbit, regardless of the initial eccentricity.
\item Eccentricity increases and coalescence due to gravitational wave emission will occur for a binary with a mass ratio of 10:1 in less than $10^8$ years ($\times M_8^{0.29}$). If the stellar distribution is triaxial or rotating the lifetime of such systems is potentially shorter.
\end{enumerate}

While we used a static, spherically symmetric background potential to account for star-star interactions, it is straightforward to extended  this method to treat more complicated cases such as an adiabatically evolving potential (e.g. due to core depletion during the BBH inspiral), non symmetric stellar models and perturbers within the scattering method.

This method can also be used to explore the time evolution of the statistical properties of the scattering events, and the extension of the dynamical friction formulation to the hard binary stage. This study will help understand the mechanism which leads to the growth of eccentricity for an $q \neq 1$ BBH, and its decay for the $q=1$ case.

\section*{Acknowledgements}
We thank David Merritt, Eugene Vasiliev, Alberto Sesana, Fazeel Mahmood Khan and Fabio Antonini for helpful discussions and comments. We thank the referee for some helpful comments.

\label{lastpage}

\end{document}